\documentclass[conference]{IEEEtran}

\pdfoutput=1

\usepackage{cite}
\usepackage{graphicx}
\graphicspath{{./}}
\usepackage{amsmath}
\usepackage{algorithmic}
\usepackage{array}
\usepackage[caption=false,font=footnotesize]{subfig}
\usepackage{url}

\usepackage{amssymb}
\usepackage{amsthm}
\usepackage{multicol}
\usepackage{multirow}
\usepackage{mdwmath}
\usepackage{mdwtab}
\usepackage{diagbox}
\usepackage{tabularx}
\usepackage{booktabs}
\usepackage{algorithm}
\usepackage{hyperref}

\usepackage{tikz}    
\usetikzlibrary{shapes,arrows,calc}

\begin{document}

\title{A Multi-Layer Blockchain Simulator and Performance Evaluation of Social Internet of Vehicles with Multi-Connectivity Management}

\author{\IEEEauthorblockN{Yi-Ting Sun, Hsin-Chieh Lee, Yun-Chen Yu, Ting-Feng Wu, Ibrahim Althamary, and Chih-Wei Huang}
    \IEEEauthorblockA{Department of Communication Engineering,
        National Central University, Taoyuan, Taiwan\\
        Email: \{111523054,112523042,113523038,113523043,106583603\}@cc.ncu.edu.tw, cwhuang@ce.ncu.edu.tw}
}


\maketitle

\begin{abstract}
    The evolution of vehicle-to-everything (V2X) communication brings significant challenges, such as data integrity and vulnerabilities stemming from centralized management. This paper presents an innovative integration of decentralized blockchain technology with V2X communication through a multi-layered architecture that combines the Simulation of Urban Mobility (SUMO) traffic simulator and the BlockSim blockchain simulator. In addition, as the Social Internet of Vehicles (SIoV) emerges, efficient resource management becomes indispensable for ensuring seamless communication. We also propose a reference multi-connectivity management method named Enhanced MAX-SINR, designed to advance research in blockchain-specific approaches, taking into account retransmission successfull rates. We evaluate blockchain performance in diverse environments such as urban, suburban, and rural areas, demonstrating that enhancing the success rate of retransmitted blockchain-related messages significantly boosts blockchain transaction performance and provides a foundation for developing intelligent SIoV systems.
\end{abstract}


\section{Introduction}

The rapid evolution of communication technology has significantly accelerated research in vehicular network communication. A major milestone in this field is vehicle-to-everything (V2X) communication, which establishes a cellular standard enhancing connectivity between vehicles and various entities\cite{noor20226g}.
The Social Internet of Vehicles (SIoV) improves V2X networks by facilitating the formation of social network-like connections among vehicles. This connectivity enables vehicles to dynamically exchange crucial information on safety, efficiency, and comfort, enhancing real-time decision-making in V2X applications.\cite{Alam2015,liu2022vrepchain}.

Blockchain technology can be adapted to improve trust, data integrity, and privacy in V2X systems\cite{rao2023blockchain}.
The decentralized nature of blockchain effectively mitigates single points of attack and failure while utilizing consensus mechanisms to verify the security of network participants and data~\cite{alladi2022comprehensive}. Building on these advantages, Liu et al.~\cite{liu2022vrepchain} introduced a blockchain-based reputation system for SIoV, designed to block the dissemination of fraudulent vehicle reputation scores among vehicles and to safeguard the privacy of these scores.

With the introduction of sidelink communication and satellite integration, transmission reliability is enhanced through sophisticated multi-connectivity management\cite{annu2024,liang2019spectrum}. Ji et al.\cite{ji2023multi} enhanced this by integrating deep reinforcement learning, which elevates spectrum utilization and adaptability to dynamic conditions. Chen et al.\cite{ipcsocial} applied role-oriented strategies in multi-agent reinforcement learning (MARL) to optimize SIoV communication efficiency, focusing on resource management tailored to vehicle social roles. Therefore, an environment where a blockchain-based application and a V2X network operate on a system with multi-connectivity deserves further study to deliver optimal performance.

In several studies, blockchain and V2X networks have been integrated and simulated through various approaches. Tu et al.\cite{tu2023secure} employed a custom simulation that allowed adjustable parameters such as transmission distance, vehicle speed, and blockchain configurations. Liu et al.\cite{liu2022vrepchain} utilized NetLogo primarily to simulate dynamic traffic scenarios, where vehicles were modeled as blockchain nodes with basic behaviors. Chavhan et al. ~\cite{chavhan2023edge} adopted the Simulation of Urban Mobility (SUMO) for simulating vehicular platoon networks to handle traffic data, while NetSim was used to simulate blockchain, emphasizing network layer activities. Nonetheless, these studies neglect to simulate a fully integrated environment that includes vehicular traffic, various connectivity options, and authentic blockchain mechanisms.

Performance analysis on blockchain is an increasingly popular topic. Gupta et al.\cite{gupta2023performance} compared consensus mechanisms in permissioned blockchains to assess their throughput, latency, and scalability under different loads. Fan et al.\cite{fan2022performance} examined transaction throughput in static nodes, adjusting blockchain parameters for testing. Kim\cite{kim2019impacts} studied the effect of node mobility on blockchain performance in V2X networks. Gao et al.\cite{gao2021multi} analyzed blockchain performance under various vehicle densities, emphasizing that transaction throughput depends on successful data transmission. Fardad et al.\cite{fardad2024blockchain} investigated blockchain performance with different connectivity options under diverse traffic loads. However, these studies have not thoroughly examined the multi-connectivity transmission mechanisms of future vehicular networks, such as the impact of satellite connections and other innovative approaches on blockchain performance.

To effectively conduct this research, we introduce a novel approach combining satellite-assisted V2X communication with blockchain technology and advanced simulation tools. This multi-layer system facilitates realistic testing and optimization of V2X networks. Leveraging the SUMO traffic simulator and the BlockSim blockchain simulator\cite{alharby2020blocksim}, alongside reference multi-connectivity management strategies, we evaluate blockchain performance in diverse environments such as urban, suburban, and rural areas.
Key contributions are summarized below:
\begin{itemize}
    \item We create a comprehensive multi-layer simulator that accurately models aspects from vehicle data generation to blockchain functionality, providing a solid foundation for future research.
    \item The implementation challenges at both the layer and system levels have been addressed to fulfill the requirements for a broad spectrum of functions.
    \item We propose a reference multi-connectivity management method that focuses on enhancing the success rate of retransmitted blockchain-related messages. This improvement significantly boosts blockchain transaction performance and provides a foundation for developing intelligent systems.
    \item The evaluation of the system architecture across various geographical regions demonstrated a significant improvement. By implementing retransmission enhancements, there was an 18.71\% increase in blockchain transaction throughput, compared to the MARL-based approach that primarily targeted non-blockchain performance objectives.
\end{itemize}
The source code of the multi-layer blockchain V2X system is also released to facilitate verification and future development\footnote{The source codes can be found at \url{https://github.com/IPCLab/V2XBlockchain.git}.}.

The paper is organized as follows: Section~\ref{sec:imple_arch} describes the V2X and blockchain-integrated system architecture.
Section~\ref{sec:bc_mechanism} discusses the blockchain mechanism and implementation challenges in V2X.
Section~\ref{sec:connectivity} covers the Multi-Connectivity Management method.
Section~\ref{sec:results} presents simulation results and performance evaluation.
Finally, Section~\ref{sec:conclusion} concludes the study and suggests future directions.

\section{The Multi-Layer System Model} \label{sec:imple_arch}

\begin{figure}
    \centering
    \includegraphics[width=1\linewidth]{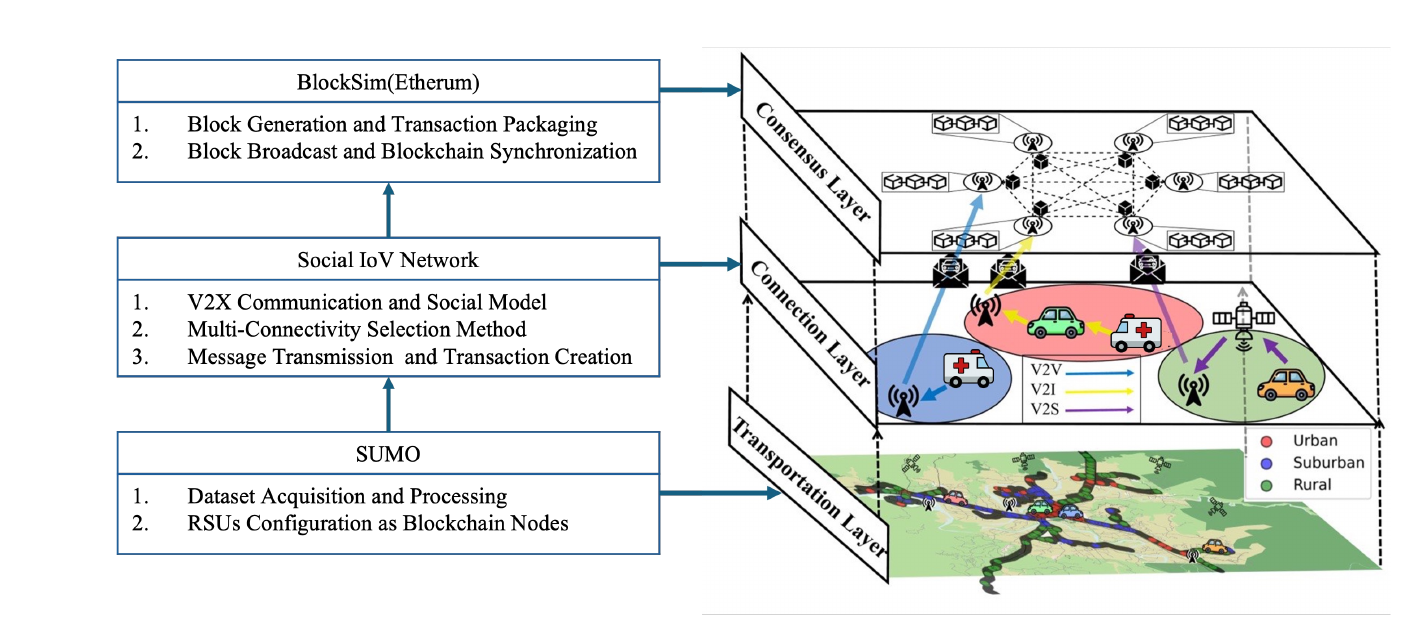}
    \caption{The the multi-layer model for integrating blockchain and SIoV networks.}
    \label{fig:Imple_Arch}
\end{figure}

Figure~\ref{fig:Imple_Arch} depicts the multi-layer integration of blockchain technology with a satellite-assisted SIoV network, highlighting the key features of each layer.

\subsection{Transportation Layer}

This layer allows customizable vehicle datasets, like those generated by SUMO, for adaptable simulations. Traffic data is categorized into urban, suburban, and rural densities, with the optional use of satellite data to enhance V2X connectivity. Road-side units (RSUs) are strategically positioned to facilitate flexible communication and function as blockchain nodes.

\subsection{Connection Layer}

The Connection Layer enables V2X communication, facilitating interactions among vehicles, infrastructure, and satellites. SIoV enhances this by creating dynamic networks for collaborative sharing of safety and traffic data, optimizing communication and resource allocation. It ensures secure message transmission across V2X types and verifies each message before packaging for blockchain transactions, enhancing data integrity and reliability.

Vehicles in the environment are denoted by the set \(\mathbf{V} = \{1, 2, \dots, v\}\), where \(v\) is the total number of vehicles. The SINR \(\gamma_{v, k}\) for a vehicle is given by:
\begin{equation}\label{eq:sinr}
    \gamma_{v, k} = \frac{PR_{v, k}}{N_0^2 + I_{v, k}^{i-1}}
\end{equation}
where \(PR_{v, k}\) is the received signal power from vehicle \(v\) on sub-channel \(k\), \(N_0^2\) is the noise variance, and \(I_{v,k}\) is the co-channel interference power from other vehicles on the same sub-channel. Each vehicle employs a multi-connectivity management approach to choose its transmission settings.

\subsection{Consensus Layer}

The Consensus Layer utilizes BlockSim\cite{alharby2020blocksim} to simulate a blockchain environment. In this setup, RSUs function as nodes that package transactions, generate blocks, and handle broadcasting and synchronization. They compete to create blocks based on their hash power, indicating their computational capability. Detailed operational mechanisms of the blockchain will be covered in the following section.

\section{Blockchain Mechanism in V2X Networks}
\label{sec:bc_mechanism}
%



In this framework, RSUs function as blockchain nodes, each equipped with a node ID, hash power, and a local blockchain initialized by a genesis block. They manage a transaction pool, converting messages into transactions and appending them to the blockchain. The following sections detail primary functions customized for transaction creation and block generation/synchronization in V2X networks.

\subsection{Message Transmission and Transaction Creation}

\begin{figure}
    \centering \includegraphics[width=1\linewidth]{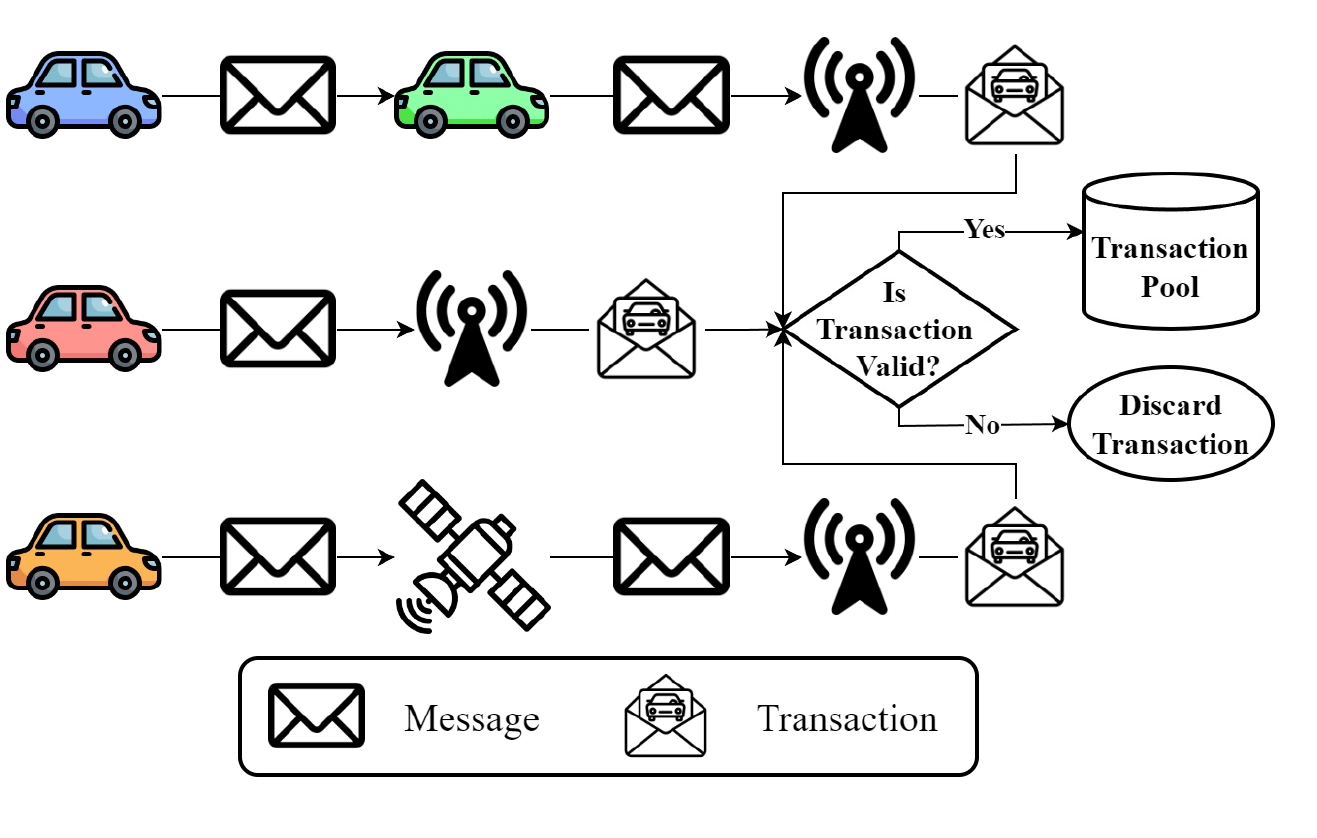}
    \caption{Message transmission and transaction creation in V2X networks.}
    \label{fig:tx_create}
\end{figure}

Figure~\ref{fig:tx_create} demonstrates the process of message transmission and transaction verification. There are three paths of transmission: vehicle-to-vehicle-to-infrastructure (V2V2I), which involves an RSU in our setup, vehicle-to-infrastructure (V2I), and vehicle-to-satellite-to-infrastructure (V2S2I). These diverse paths ensure that messages can reach a blockchain node even in the absence of a nearby RSU. The message latency $L_M$ is defined as the time it takes for a message to travel from the vehicle to the RSU. Upon receiving a message, the RSU verifies its integrity and then encapsulates it as a transaction for further processing within the blockchain network. Furthermore, a Gaussian distribution is used to allocate the hash power for each RSU miner.

\subsection{Block Generation and Transaction Packaging}\label{ssec:bc_arch}

\begin{figure}
    \centering
    \includegraphics[width=0.8\linewidth]{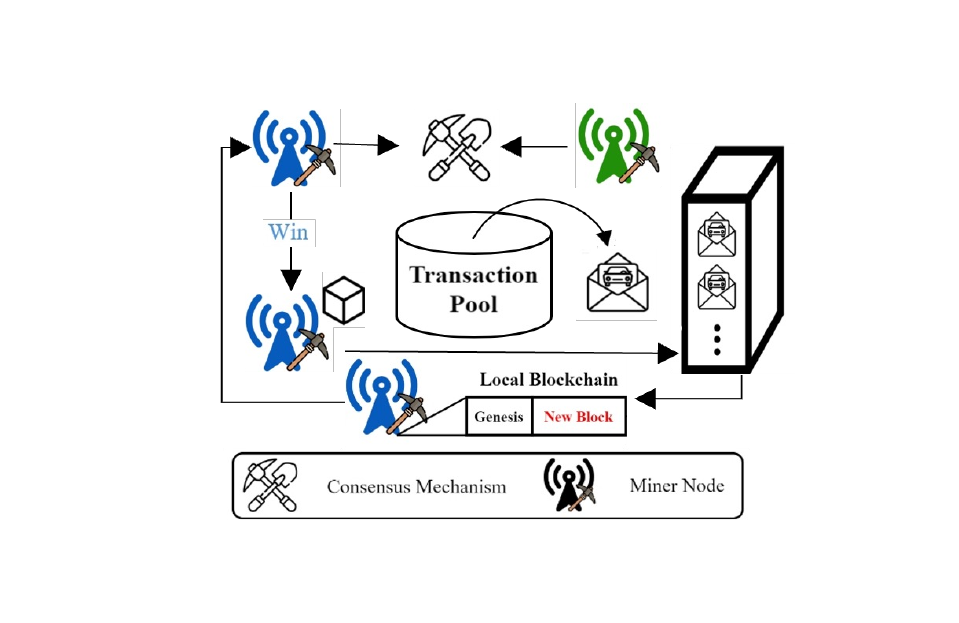}
    \caption{Block Generation and Transaction Packaging Process.}
    \label{fig:block_generate}
\end{figure}

The block generation process is illustrated in \figurename~\ref{fig:block_generate}. Throughout the simulation, miner nodes actively engage in the consensus mechanism. They compete to generate blocks, with the fastest miner considered the winner. The winning node selects transactions from the pool to add to its local blockchain. After confirming the block is stored locally, the miner re-enters the consensus process for the next round. The block generation time $t_G$ is modeled by an exponential distribution:
\begin{equation}
    t_G \sim \text{Exponential}(\lambda = R_n \times \frac{1}{T_{G}} )
\end{equation}
where ${R_n}$ denotes the hash power value $H_n$ of the miner node $n$ as a proportion of all miner nodes. ${T_{G}}$ represents the average block generation time, which is a constant value.

\subsection{Block Broadcast and Blockchain Synchronization}

The block broadcast process, essential for network synchronization, is shown in green in \figurename~\ref{fig:timeline}. Miners broadcast the generated block across the network, with the receiving time of other nodes $t_R$ following an exponential distribution:
\begin{equation}
    t_R \sim \text{Exponential}(\lambda = \frac{1}{T_{R}} )
\end{equation}
where ${t_R}$ is a constant, represents the average time for a node to receive the block. Once a receiving node gets the broadcasted block, it will validate the block. If the block’s \emph{Hash Value}~\cite{meijers2022blockchain}, which uniquely links each block to its predecessor, doesn’t match the last block in the receiving node’s blockchain, the block is discarded to ensure continuity and data integrity.

\subsection{Implementation Challenges and Solutions}

Integrating blockchain with V2X (Vehicle-to-Everything) systems presents challenges. Layer-specific issues can be addressed within individual layers, whereas system-wide challenges affect the whole system.

\subsubsection{Layer-specific issues}

To balance high data rates with security and decentralization in V2X and blockchain integration, propose that vehicles upload only essential Key Messages—vehicle ID, speed, and location—to reduce bandwidth and storage. Introduce a verification method between the Connection and Consensus Layers, storing only fully transmitted messages in the transaction pool to ensure integrity.

\subsubsection{System-wide issues}

\begin{figure}
    \centering
    \includegraphics[width=1\linewidth]{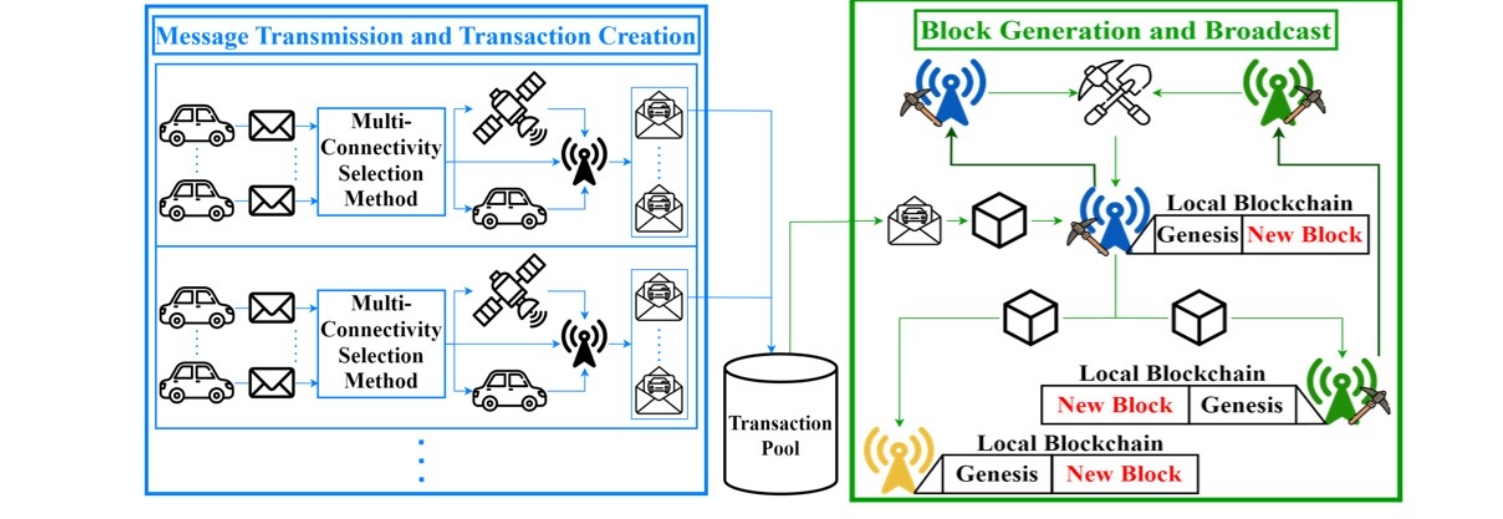}
    \caption{Integrated System for Parallel Processing of Two Timelines. The blue and green sections are two timelines respectively.}
    \label{fig:timeline}
\end{figure}

Simulating parallel multitasking poses a significant challenge in system integration. During the implementation of the three-layer architecture shown in \figurename~\ref{fig:Imple_Arch}, it was observed that the system operates on two timelines as illustrated in Figure \figurename~\ref{fig:timeline}. This necessitates the use of a global clock for synchronization. Message latency to the RSU is determined by the connection mode, affecting simulation time. When the global clock matches the simulated arrival time at the RSU, the message is verified and added to the transaction pool. Similarly, at the designated block generation time, miners bundle transactions into a block, broadcasting it to receiving nodes as the clock aligns with the reception time. The global clock maintains the correct event execution sequence.

\section{Multi-Connectivity Management}\label{sec:connectivity}
%


We introduce an improved multi-connectivity management technique named Enhanced MAX-SINR as reference, designed to advance research in blockchain-specific approaches. This method builds upon the conventional MAX-SINR technique. Given that message retransmission poses a significant challenge to blockchain transaction throughput\cite{fardad2024blockchain}, Enhanced MAX-SINR dynamically switches channels and connections to mitigate congestion, thus expediting message delivery forming successful transactions.

\begin{algorithm}
    \caption{Enhanced MAX-SINR Selection Method}
    \label{alg:E-MAXSINR}
    \begin{algorithmic}[1]
        \REQUIRE
        $i$: the message transmitted for the $i$-th time;
        $\mathbf{V}$: the set of each vehicle $v$ in the current environment;
        ${\mathbf{A}_v}^{(i-1)}$: the set of previous transmission selections for each vehicle $v$, including connection mode $M_v^{(i-1)}$, transmission signal power $PT_v^{(i-1)}$, sub-channel $k_v^{(i-1)}$, and transmission success or failure ${SF}_v^{(i-1)}$;
        \ENSURE
        $\mathcal{A}^{i}$: the set of transmission selections for the $i$-th transmission of all vehicles
        \small
        \STATE Each vehicle selects an action for this transmission
        \FOR{each vehicle $v \in \mathbf{V}$}
        \IF{$i = 0$}
        \STATE ${\mathbf{A}_v}^{i} \gets$ MAX-SINR selection method
        \ELSIF{$SF_v^{(i-1)} = \text{Failure}$}
        \STATE $strategy = \text{random integer in } \{0, 1, 2, 3\}$

        \STATE \textbf{switch} $strategy$
        \STATE \hspace{1em} \textbf{case 0}: \begin{minipage}[t]{0.77\linewidth}
            Select ${\mathbf{A}_v}^{i}$ with the maximum SINR in a connection mode other than $M_v^{(i-1)}$
        \end{minipage}%
        \STATE \hspace{1em} \textbf{case 1}: \begin{minipage}[t]{0.77\linewidth}
            Select ${\mathbf{A}_v}^{i}$ with the maximum SINR in the same connection mode, but with a sub-channel other than $k_v^{(i-1)}$
        \end{minipage}%
        \STATE \hspace{1em} \textbf{case 2}: \begin{minipage}[t]{0.77\linewidth}
            ${\mathbf{A}_v}^{i} \gets {\mathbf{A}_v}^{(i-1)}$
        \end{minipage}%
        \STATE \hspace{1em} \textbf{case 3}: \begin{minipage}[t]{0.77\linewidth}
            ${\mathbf{A}_v}^{i} \gets$ MAX-SINR selection method
        \end{minipage}%
        \STATE \textbf{end switch}
        \ENDIF
        \ENDFOR
        \RETURN $\mathcal{A}^{i}$
        \normalsize
    \end{algorithmic}
\end{algorithm}

The action determined by the vehicle $v$ as a smart agent is denoted by the set $\textbf{A}_v$ and the collective actions of all vehicles are represented by $\mathcal{A}$ :
\begin{equation}
    \textbf{A}_v = \{M_v, PT_v, k_v\}, \quad
    \mathcal{A} = \{A_1, A_2, \dots, A_v\}
\end{equation}
where \( M_v \) denotes the chosen connection mode, \( PT_v \) is the transmission signal power, and \( k_v \) refers to the sub-channel.
When the MAX-SINR method attempts a retransmission after a failure, it considers only interference from the previous transmission to calculate SINR, limiting the vehicle's ability to avoid congested channels and risking further failures. To address this, The Enhanced MAX-SINR method in Algorithm~\ref{alg:E-MAXSINR}, offering four additional selection strategies for vehicles, based on comparison with the previous transmission:
\begin{itemize}
    \item Change Connection Mode: Select a channel with the maximum SINR under a different connection mode.
    \item Change Sub-channel: Select a different sub-channel with the maximum SINR under the same connection mode.
    \item Keep transmission selection: Use the same action.
    \item Reapply MAX-SINR: Reapply the MAX-SINR method to determine a new action.
\end{itemize}

\section{Performance Evaluation}\label{sec:results}

\subsection{Simulation Setup}

\begin{figure}
    \centering
    \includegraphics[width=0.9\linewidth]{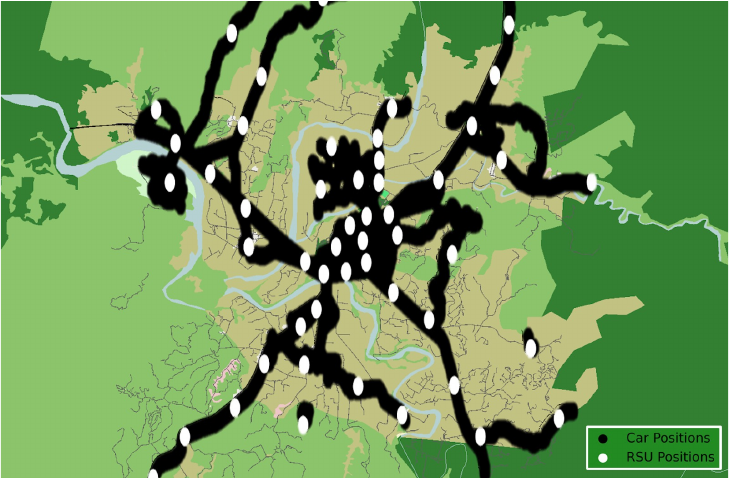}
    \caption{Vehicles and RSUs Distribution Generated by SUMO Simulation. }
    \label{fig:map}
\end{figure}

We use SUMO to simulate traffic flows in the Yujing District of Tainan City. The area offers a diverse mix of urban, suburban, and rural regions, with vehicle densities of over 2000, 1000, and 400 vehicles per km respectively, according to Taiwan’s Ministry of Transportation. In a 50-second simulation using SUMO, average vehicle counts of 100 to 500 result in densities of approximately 8.47 to 42.34 vehicles per km² across a map of $8.7 \times 11.7$ km². The actual road network spans 11.81 km². Satellite data from CelesTrak\cite{celestrak_starlink_data} enhances the V2X environment, and RSU deployment is shown in \figurename~\ref{fig:map}. Details on simulation parameters are provided in Table \ref{table:sim_parameters}. The simulation runs on a PC with an Intel Core i7-13700 CPU and 32 GB memory.

\begin{table}
    \renewcommand{\arraystretch}{1.2}
    \caption{Simulation Parameters}
    \label{table:sim_parameters}
    \centering
    \scalebox{0.8}{%
        \begin{tabular}{|p{4.5cm}|p{5.5cm}|}
            \hline
            \textbf{Parameter}                    & \textbf{Value}                                                        \\
            \hline
            Area (km$^2$)                         & 8.7 $\times$ 11.7                                                     \\
            \hline
            RSUs(Miner; Regular)                  & 54(30; 24)                                                            \\
            \hline
            RSU Coverage Distance (m)             & 500                                                                   \\
            \hline
            RAU Placement Distance (m)            & urban: 450; suburban: 750; rural: 1000                                \\
            \hline
            Carrier Frequency (GHz)               & 3.5                                                                   \\
            \hline
            Antenna (Height, Gain)                & Vehicle: 1.5 m, 3 dBi; BS: 25 m, 8 dBi                                \\
            \hline
            Noise Figure (dB)                     & Vehicle: 9; BS: 5; Satellite: 1.2                                     \\
            \hline
            Satellite (Orbit, Altitude, Band)     & LEO, 550 km, Ka (30 GHz)                                              \\
            \hline
            Sub-channels (MHz)                    & V2V\&V2I: $5 \times 1$; V2S: $10 \times 20$                           \\
            \hline
            Message Payload Size (bytes)          & 650                                                                   \\
            \hline
            Satellite Tx/Rx Gain                  & 43.2 dBm / 30.5 dBi                                                   \\
            \hline
            Transmit Power (dBm)                  & V2V: [23, 10, 15, 17]; V2I: 23; V2S: 33.5                             \\
            \hline
            Thermal Noise Floor (dBm)             & -174                                                                  \\
            \hline
            Scintillation Loss (dB)               & 2.2                                                                   \\
            \hline
            Fading                                & Shadowing: Log-normal; Fast: Rayleigh                                 \\
            \hline
            Pathloss Model                        & V2I,V2V: TR 38.886~\cite{3gpp38.886} ; V2S: TR 38.821~\cite{TR38.821} \\
            \hline
            $simTime$: Simulation Time (s)        & 50                                                                    \\
            \hline
            Key Message Generation Period (s)     & 0.5                                                                   \\
            \hline
            Maximum Transmission Delay (ms)       & 3                                                                     \\
            \hline
            Hash Power Distribution               & Gaussian Distribution                                                 \\
            \hline
            $T_G$: Block Interval (s)             & 2.7~\cite{son2021effective}                                           \\
            \hline
            $T_R$: Average Block Receive Time (s) & 0.25~\cite{son2021effective}                                          \\
            \hline
            ${GL}_B$: Block Gas Limit (s)         & 30,000,000 \href{https://etherscan.io/}{Etherscan}                    \\
            \hline
        \end{tabular}}
\end{table}


To assess the effectiveness of integrating V2X with blockchain networks, we employ the following performance metrics.
\begin{itemize}
    \item Message throughput: This metric measures the system-wide number of messages that are successfully transmitted from vehicles to Roadside Units (RSUs) per millisecond and subsequently become valid transactions.
          \begin{equation}
              \text{Message Throughput (M/ms)} = \frac{ME}{\sum {L_M}},
          \end{equation}
          where $ME$ represents the total count of messages that were successfully transmitted and validated as transactions. The term $\sum L_M$ indicates the cumulative latency experienced by all messages to enter the transaction pool.
    \item Transaction throughput: This metric quantifies the average number of transactions that are uploaded to the blockchain every second, which is crucial for evaluating the performance of blockchain applications\cite{fan2020performance}.
          \begin{equation}\label{eq:tx_through}
              \text{Transaction Throughput (TX/s)} = \frac{\sum TX}{simTime},
          \end{equation}
          where $\sum TX$ denotes the total count of transactions successfully recorded on the blockchain after applying the longest chain principle, and $simTime$ represents the total simulation duration.
\end{itemize}



The performance evaluation includes MARL\cite{ipcsocial}, MAX-SINR, and Random methods, along with the reference multi-connectivity management design Enhanced MAX-SINR. The MARL approach allows vehicles to optimize a utility function aimed at maximizing performance targets for non-blockchain applications, taking into account throughput and delay constraints. Here, each vehicle functions as an independent agent that selects actions based on its current state, which includes connection mode, transmission power, and sub-channel selection. The MARL method employs a global reward system, aggregating immediate rewards of each vehicle's transmission performance at any given moment. The MAX-SINR method optimizes connection by selecting the mode and sub-channel with maximum SINR, as per \eqref{eq:sinr}. In contrast, the Random method uses a uniform distribution for vehicle transmission selection.

\subsection{Performance Evaluation Results}

\begin{figure}
    \centering
    \includegraphics[width=0.9\linewidth]{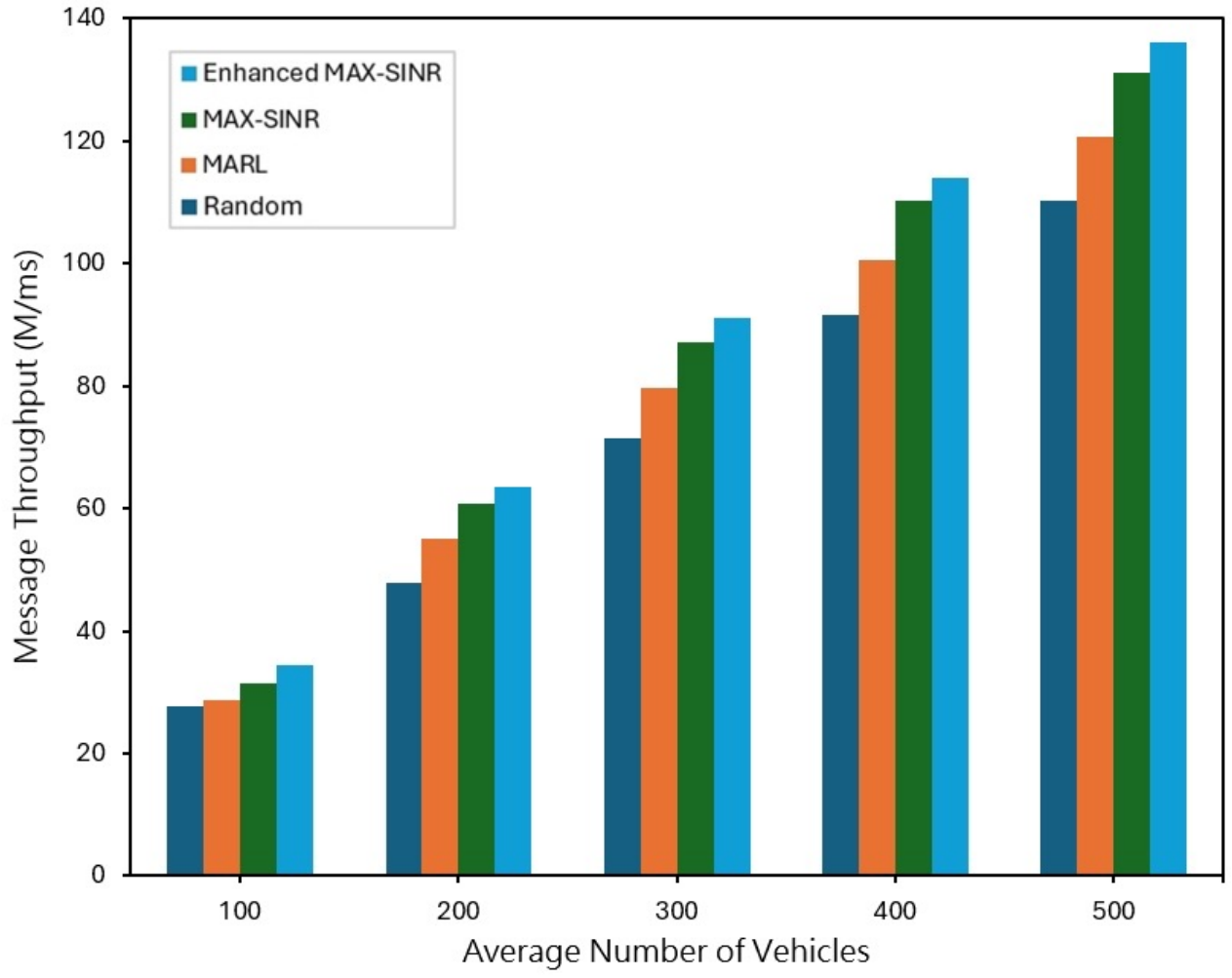}
    \caption{Message throughput under different average numbers of vehicles }
    \label{fig:me_throughput}
\end{figure}

Message throughput performance across four multi-connectivity selection methods, as shown in~\figurename~\ref{fig:me_throughput}. While the vehicle numbers rise, all methods exhibit an upward trend in key message generation. Our reference Enhanced MAX-SINR method consistently outperforms others due to its additional selection strategies, which boost successful data transmission rates, especially in congested environments. Notably, in a V2X environment with 500 vehicles, our reference Enhanced MAX-SINR method achieves a message throughput of 135.95 M/ms. The MAX-SINR method follows with 131.06 M/ms but faces real-time congestion issues, placing it second. The MARL method optimizes channel resource use by balancing throughput, latency, and power, achieving 120.64 M/ms. It slightly surpasses the Random method, which performs poorly due to ineffective option selection in varying network conditions.

\begin{figure}
    \centering
    \includegraphics[width=0.9\linewidth]{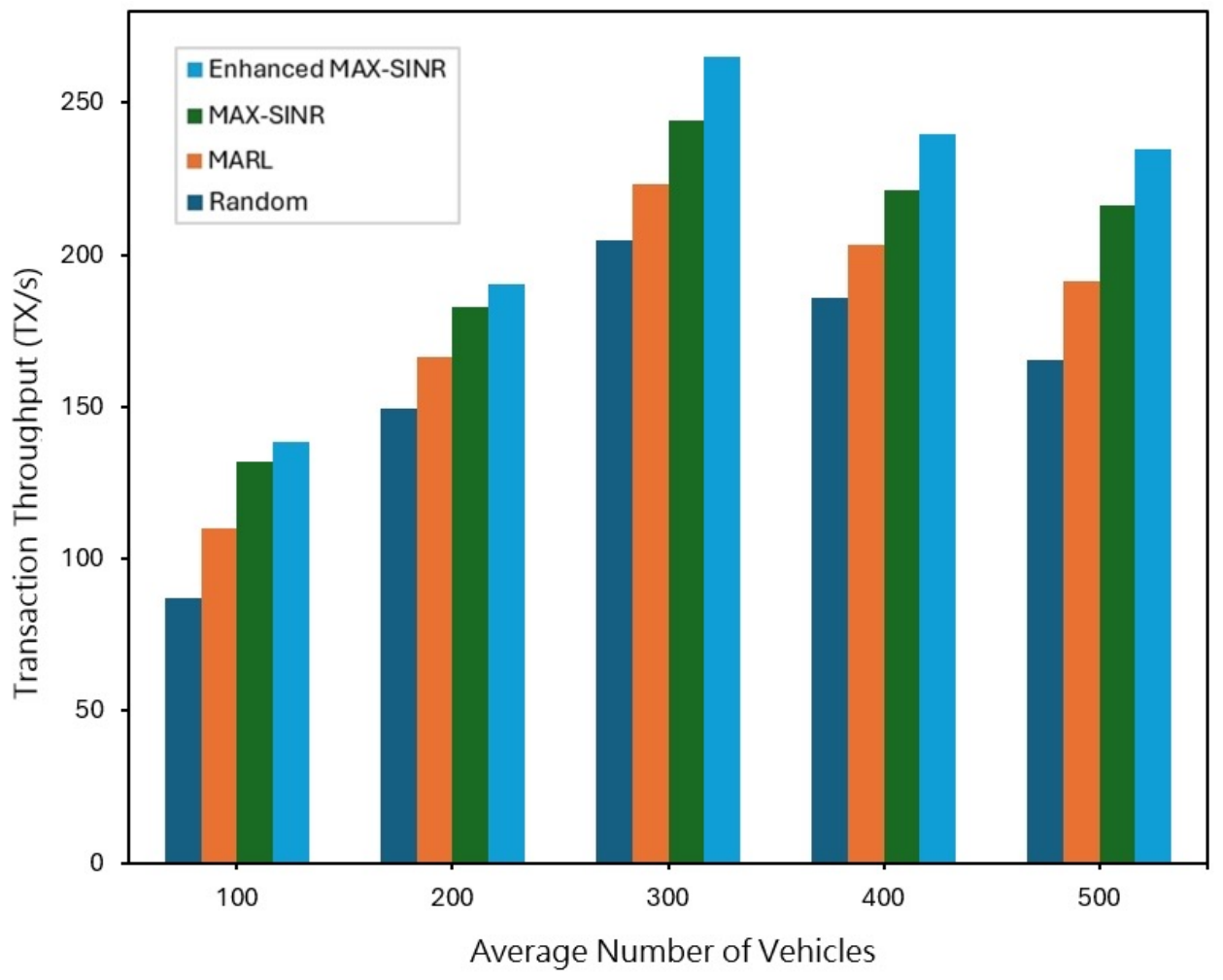}
    \caption{Transaction throughput under different average numbers of vehicles }
    \label{fig:tx_throughput}
\end{figure}

\figurename~\ref{fig:tx_throughput} shows transaction throughput for various multi-connectivity selection methods. The Enhanced MAX-SINR method consistently outperforms others in terms of transaction throughput, achieving 264.286 TX/s at 300 vehicles—an 18.71\% increase over MARL’s 222.62 TX/s. Though MARL optimizes resource utilization, it is not tailored for blockchain, leading to lower throughput. MAX-SINR also faces channel congestion issues, causing transmission failures, and the Random method performs worst due to a lack of congestion management. As vehicle numbers near 300, throughput caps across methods, revealing blockchain’s transaction limits at high densities and how network conditions affect performance.

\begin{figure}
    \centering
    \includegraphics[width=0.95\linewidth]{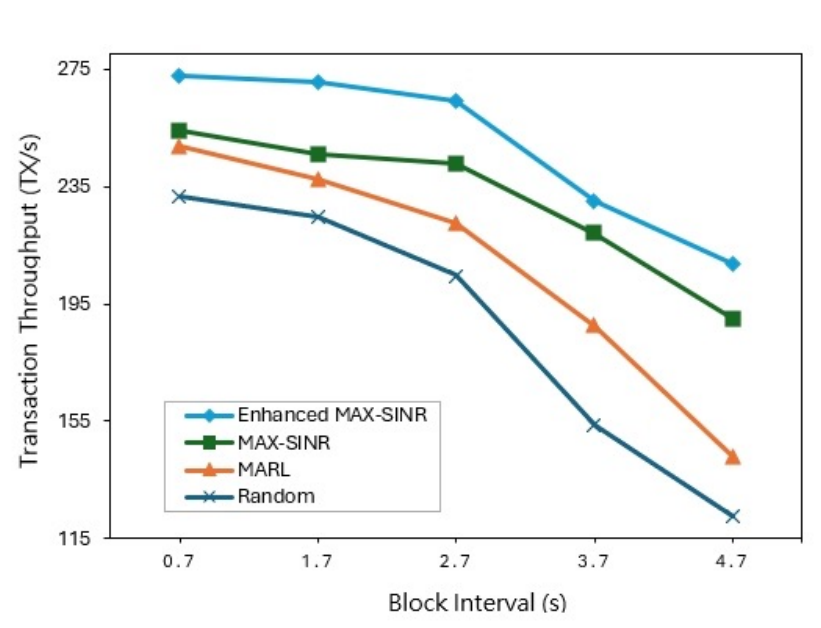}
    \caption{Comparison of four methods under different block creation speeds }
    \label{fig:Block_Interval}
\end{figure}

\figurename~\ref{fig:Block_Interval} shows how adjusting block creation time in a V2X environment with 300 vehicles impacts blockchain performance. Shorter block times may increase throughput but also raise the risk of chain forks, leading to unconfirmed transactions and reduced throughput. Balancing block generation speed and fork risk is essential, especially under high traffic. Despite these challenges, the Enhanced MAX-SINR method consistently achieves the highest throughput at 272.548 TX/s, showing adaptability across varying block intervals.

The experiments emphasize the need to redesign multi-connectivity selection methods in integrated blockchain and V2X networks to enhance blockchain processing performance, considering the impact of the V2X environment.

\section{Conclusion}\label{sec:conclusion}

We presented a multi-layer blockchain simulator for satellite-assisted SIoV scenarios, resolving key implementation issues. Performance evaluations across varying traffic volumes underscore the importance of multi-connectivity management design for optimal blockchain performance. Future advancements could involve developing and testing intelligent management methods using the open-source simulator.


\bibliographystyle{IEEEtran}
\bibliography{V2XBlockchain_arXiv}

\begin{thebibliography}{10}
\providecommand{\url}[1]{#1}
\csname url@samestyle\endcsname
\providecommand{\newblock}{\relax}
\providecommand{\bibinfo}[2]{#2}
\providecommand{\BIBentrySTDinterwordspacing}{\spaceskip=0pt\relax}
\providecommand{\BIBentryALTinterwordstretchfactor}{4}
\providecommand{\BIBentryALTinterwordspacing}{\spaceskip=\fontdimen2\font plus
\BIBentryALTinterwordstretchfactor\fontdimen3\font minus
  \fontdimen4\font\relax}
\providecommand{\BIBforeignlanguage}[2]{{%
\expandafter\ifx\csname l@#1\endcsname\relax
\typeout{** WARNING: IEEEtran.bst: No hyphenation pattern has been}%
\typeout{** loaded for the language `#1'. Using the pattern for}%
\typeout{** the default language instead.}%
\else
\language=\csname l@#1\endcsname
\fi
#2}}
\providecommand{\BIBdecl}{\relax}
\BIBdecl

\bibitem{noor20226g}
M.~Noor-A-Rahim, Z.~Liu, H.~Lee, M.~O. Khyam, J.~He, D.~Pesch, K.~Moessner,
  W.~Saad, and H.~V. Poor, ``{6G for vehicle-to-everything (V2X)
  communications: Enabling technologies, challenges, and opportunities},''
  \emph{Proceedings of the IEEE}, vol. 110, no.~6, pp. 712--734, 2022.

\bibitem{Alam2015}
K.~M. Alam, M.~Saini, and A.~E. Saddik, ``{Toward Social Internet of Vehicles:
  Concept, Architecture, and Applications},'' \emph{IEEE Access}, vol.~3, pp.
  343--357, 2015.

\bibitem{liu2022vrepchain}
Y.~Liu, Z.~Xiong, Q.~Hu, D.~Niyato, J.~Zhang, C.~Miao, C.~Leung, and Z.~Tian,
  ``{Vrepchain: A decentralized and privacy-preserving reputation system for
  social internet of vehicles based on blockchain},'' \emph{IEEE Transactions
  on Vehicular Technology}, vol.~71, no.~12, pp. 13\,242--13\,253, 2022.

\bibitem{rao2023blockchain}
P.~M. Rao, S.~Jangirala, S.~Pedada, A.~K. Das, and Y.~Park, ``{Blockchain
  integration for IoT-enabled V2X communications: a comprehensive survey,
  security issues and challenges},'' \emph{IEEE Access}, vol.~11, pp.
  54\,476--54\,494, 2023.

\bibitem{alladi2022comprehensive}
T.~Alladi, V.~Chamola, N.~Sahu, V.~Venkatesh, A.~Goyal, and M.~Guizani, ``{A
  comprehensive survey on the applications of blockchain for securing vehicular
  networks},'' \emph{IEEE Communications Surveys \& Tutorials}, vol.~24, no.~2,
  pp. 1212--1239, 2022.

\bibitem{annu2024}
Annu and P.~Rajalakshmi, ``{Towards 6G V2X Sidelink: Survey of Resource
  Allocation—Mathematical Formulations, Challenges, and Proposed
  Solutions},'' \emph{IEEE Open Journal of Vehicular Technology}, vol.~5, pp.
  344--383, 2024.

\bibitem{liang2019spectrum}
L.~Liang, H.~Ye, and G.~Y. Li, ``{Spectrum Sharing in Vehicular Networks Based
  on Multi-Agent Reinforcement Learning},'' \emph{IEEE Journal on Selected
  Areas in Communications}, vol.~37, no.~10, pp. 2282--2292, 2019.

\bibitem{ji2023multi}
Y.~Ji, Y.~Wang, H.~Zhao, G.~Gui, H.~Gacanin, H.~Sari, and F.~Adachi,
  ``{Multi-Agent Reinforcement Learning Resources Allocation Method Using
  Dueling Double Deep Q-Network in Vehicular Networks},'' \emph{IEEE
  Transactions on Vehicular Technology}, 2023.

\bibitem{ipcsocial}
P.-Y. Chen, Y.-H. Zheng, I.~Althamary, J.-L. Chern, and C.-W. Huang,
  ``{Multi-Agent Deep Reinforcement Learning for Spectrum Management in V2X
  with Social Roles},'' in \emph{GLOBECOM 2023-2023 IEEE Global Communications
  Conference}.\hskip 1em plus 0.5em minus 0.4em\relax IEEE, 2023, pp.
  2293--2298.

\bibitem{tu2023secure}
S.~Tu, H.~Yu, A.~Badshah, M.~Waqas, Z.~Halim, and I.~Ahmad, ``{Secure Internet
  of Vehicles (IoV) with decentralized consensus blockchain mechanism},''
  \emph{IEEE Transactions on Vehicular Technology}, vol.~72, no.~9, pp.
  11\,227--11\,236, 2023.

\bibitem{chavhan2023edge}
S.~Chavhan, S.~Kumar, P.~Tiwari, X.~Liang, I.~H. Lee, and K.~Muhammad,
  ``{Edge-enabled Blockchain-based V2X Scheme for Secure Communication within
  the Smart City Development},'' \emph{IEEE Internet of Things Journal}, 2023.

\bibitem{gupta2023performance}
M.~K. Gupta, R.~K. Dwivedi, A.~Sharma, M.~Farooq \emph{et~al.}, ``{Performance
  Evaluation of Blockchain Platforms},'' in \emph{2023 International Conference
  on IoT, Communication and Automation Technology (ICICAT)}.\hskip 1em plus
  0.5em minus 0.4em\relax IEEE, 2023, pp. 1--6.

\bibitem{fan2022performance}
C.~Fan, C.~Lin, H.~Khazaei, and P.~Musilek, ``{Performance analysis of
  hyperledger besu in private blockchain},'' in \emph{2022 IEEE international
  conference on decentralized applications and infrastructures (DAPPS)}.\hskip
  1em plus 0.5em minus 0.4em\relax IEEE, 2022, pp. 64--73.

\bibitem{kim2019impacts}
S.~Kim, ``{Impacts of mobility on performance of blockchain in VANET},''
  \emph{IEEE Access}, vol.~7, pp. 68\,646--68\,655, 2019.

\bibitem{gao2021multi}
L.~Gao, C.~Wu, T.~Yoshinaga, X.~Chen, and Y.~Ji, ``{Multi-channel blockchain
  scheme for internet of vehicles},'' \emph{IEEE Open Journal of the Computer
  Society}, vol.~2, pp. 192--203, 2021.

\bibitem{fardad2024blockchain}
M.~Fardad, G.-M. Muntean, and I.~Tal, ``{A Blockchain-Enabled Vehicular Edge
  Computing Framework for Secure Performance-oriented V2X Service Delivery},''
  \emph{IEEE Transactions on Vehicular Technology}, 2024.

\bibitem{alharby2020blocksim}
M.~Alharby and A.~van Moorsel, ``{Blocksim: An extensible simulation tool for
  blockchain systems},'' \emph{Frontiers in Blockchain}, vol.~3, p.~28, 2020.

\bibitem{meijers2022blockchain}
J.~Meijers, P.~Michalopoulos, S.~Motepalli, G.~Zhang, S.~Zhang, A.~Veneris, and
  H.-A. Jacobsen, ``{Blockchain for v2x: Applications and architectures},''
  \emph{IEEE Open Journal of Vehicular Technology}, vol.~3, pp. 193--209, 2022.

\bibitem{celestrak_starlink_data}
{North American Aerospace Defense Command (NORAD)}, ``{Starlink Satellite
  Elements},'' \url{http://celestrak.com/NORAD/elements/}, 2023, accessed:
  August 2023.

\bibitem{3gpp38.886}
3GPP, ``{V2X Services based on NR; User Equipment (UE) radio transmission and
  reception; (Release 16)},'' 3GPP, Tech. Rep., March 2021.

\bibitem{TR38.821}
------, ``{3GPP TR 38.821: Solutions for NR to support non-terrestrial networks
  (NTN)},'' 3GPP, Tech. Rep., April 2023.

\bibitem{son2021effective}
D.~H. Son, T.~T.~T. Quynh, T.~V. Khoa, D.~T. Hoang, N.~L. Trung, N.~V. Ha,
  D.~Niyato, D.~N. Nguyen, and E.~Dutkiewicz, ``{An Effective Framework of
  Private Ethereum Blockchain Networks for Smart Grid},'' in \emph{2021
  International Conference on Advanced Technologies for Communications
  (ATC)}.\hskip 1em plus 0.5em minus 0.4em\relax IEEE, 2021, pp. 312--317.

\bibitem{fan2020performance}
C.~Fan, S.~Ghaemi, H.~Khazaei, and P.~Musilek, ``{Performance evaluation of
  blockchain systems: A systematic survey},'' \emph{IEEE Access}, vol.~8, pp.
  126\,927--126\,950, 2020.

\end{thebibliography}

\end{document}